\documentclass[a4paper,fleqn]{cas-dc}

\usepackage[numbers,sort&compress]{natbib}

\def\tsc#1{\csdef{#1}{\textsc{\lowercase{#1}}\xspace}}
\tsc{WGM}
\tsc{QE}
\tsc{EP}
\tsc{PMS}
\tsc{BEC}
\tsc{DE}

\begin{document}
\let\WriteBookmarks\relax
\def\floatpagepagefraction{1}
\def\textpagefraction{.001}
\shorttitle{Sn$_{0.06}$Cr$_3$Te$_4$: A Skyrmion Superconductor}
\shortauthors{Shubham et~al.}
\title [mode = title]{Sn$_{0.06}$Cr$_3$Te$_4$: A Skyrmion Superconductor}
\author[1]{Shubham Purwar}
\credit{Conceptualization, Methodology, Validation, Formal analysis, Investigation, Writing original draft, Writing review \& editing}
\author[2]{Anumita Bose}
\credit{DFT calculations, Writing review \& editing}
\author[1]{Achintya Low}
\credit{Data curation, Writing review \& editing}
\author[3]{Satyendra Singh}
\credit{Data curation, Heat capacity measurements}
\author[3]{R. Venkatesh}
\credit{Data curation, Heat capacity measurements, Writing review \& editing}
\author[2]{Awadhesh Narayan}[orcid=0000-0003-0127-7047]
\credit{Conceptualization, DFT Calculations, Validation, Resources, Writing review \& editing}
\ead{awadhesh@iisc.ac.in}
\author[1] {Setti Thirupathaiah}[orcid=0000-0003-1258-0981]
\ead{setti@bose.res.in}
\ead[URL]{www.qmat.in}
\credit{Conceptualization, Validation, Resources, Supervision, Project administration, Funding acquisition, Writing review \&
editing}

\affiliation[1]{Department of Condensed Matter and Materials Physics, S. N. Bose National Centre for Basic Sciences, Kolkata, West Bengal 700106, India.}
\affiliation[2]{Solid State and Structural Chemistry Unit, Indian Institute of Science, Bengaluru 560012, India.}
\affiliation[3]{Low-Temperature Laboratory, UGC-DAE Consortium for Scientific Research, Indore, Madhya Pradesh 452017, India.}

\begin{abstract}
Topological superconductors are an exciting class of quantum materials from the point of view of the fundamental sciences and potential technological applications. Here, we report on the successful introduction of superconductivity in a ferromagnetic layered skyrmion system Cr$_3$Te$_4$, obtained by the Sn intercalation, below a transition temperature of $T_c$$\approx$3.5 K. We observe several interesting physical properties, such as superconductivity, magnetism, and the topological Hall effect, simultaneously in this system. Despite the magnetism and Meissner effects being anisotropic, the superconductivity observed from the in-plane electrical resistivity ($\rho_{\it{bb}}$)  is nearly isotropic between $H\parallel \it{bc}$ and $H\parallel \it{a}$, suggesting separate channels of conduction electrons responsible for the superconductivity and magnetism of this system, which is also supported by our spin-resolved DFT calculations. We identify two orders of higher carrier density in superconducting Sn$_{0.06}$Cr$_{3}$Te$_4$ than the parent Cr$_3$Te$_4$. A jump in the specific heat is noticed around the $T_c$ with a volume fraction of 33\%, confirming the bulk superconductivity in Sn$_{0.06}$Cr$_{3}$Te$_4$. In addition to the introduction of superconductivity, tuning of topological Hall properties is noticed with Sn intercalation. Our observation of superconductivity in a skyrmion lattice brings up a new class of topological quantum materials.
\end{abstract}


\maketitle

\section{Introduction}
Superconductivity observed with a topological quantum phase is a much-anticipated phenomenon as it opens a new window of potential technological applications in superconducting electronics, such as topological quantum computations~\cite{Nayak2008}. Thus, designing new materials and their experimental realization is an essential scientific sequence for garnering futuristic quantum technological applications. Several proposals exist for achieving topological superconductivity, such as the heterostructure of a superconductor and a topological insulator or a doped topological insulator. In topological superconductors, the Cooper pairs are trapped at the vertices of the interface in the form of Majorana fermions due to the proximity effect~\cite{Fu2008, Nilsson2008, Veldhorst2012, Scammell2022} or at the junction of a superconductor and a magnet deposited on the surface of a topological insulator~\cite{Fu2009}. Nevertheless, several experimental studies were carried out to realize the superconductivity in the presence of quantum topology~\cite{Akhmerov2009, Wray2010, Banerjee2018}.

In addition, superconductivity has been found in many van der Waals (vdW) materials such as NaSn$_2$As$_2$ ($T_c=1.3$ K)~\cite{goto2017snas}, Na$_{1-x}$Sn$_2$P$_2$ ($T_c=2$ K)~\cite{goto2018na1}, NbSe$_2$ ($T_c=6$ K)~\cite{xiaoxiang2015strongly}, and Re$_6$Se$_8$Cl$_2$ ($T_c=8$ K)~\cite{telford2020doping}. Further, there exists another class of vdW materials in which the superconductivity is induced by the metal-ion intercalation, such as A$_x$Bi$_2$Se$_3$ (A = Sr, Cu)~\cite{PhysRevB.92.020506, Hor2010}, Sn$_{0.5}$TaSe$_2$ \cite{adam2021superconducting}, Sn$_x$NbSe$_2$ \cite{naik2019effect}, and Cu$_x$TiSe$_2$~\cite{Morosan2006}. In these systems, the intercalated metal ions act as charge carrier donors, increasing the carrier density near the Fermi level and, thus, the superconductivity. Most importantly, all these intercalated vdW superconductors are nonmagnetic and do not have chiral spin structure in their pristine phase.

On the other hand, the layered Cr$_x$Te$_y$ based systems are potential candidates to have chiral spin structures originating the topological skyrmion lattice~\cite{Liu2022, Saha2022, Li2022a, Purwar2023, Fujisawa2023, Chi2023, Zhang2023}. Notably, these systems are formed by the alternative stacking of Cr-full (CrTe$_2$-layer) and Cr-vacant (intercalated Cr-layer) layers along the crystal growth axis~\cite{Dijkstra1989}. Therefore, the intercalated Cr concentration is critical in forming the magnetic and topological properties. As the van der Waals gap separates CrTe$_2$ layers, introducing the alien atoms within the vdW gap is quite feasible. We choose to intercalate Sn within the vdW gap of CrTe$_2$-layers to induce the superconductivity in a topological ferromagnet, Cr$_3$Te$_4$~\cite{Purwar2023}. This way, the intriguing exotic quantum phases, ferromagnetism, and superconductivity can be achieved in a topological skyrmion lattice. It is worth mentioning here that no study is available to date demonstrating the superconducting phase within a skyrmion lattice. Further, such an experimental observation is quite challenging as the ground state properties like ferromagnetism and superconductivity are found to be competing, if not coexisting, unless the ferromagnetism is weak~\cite{Pfleiderer2001, Yang2005, Huy2007}.

In this study, we successfully intercalated Sn within the van der Waals gap of topological Cr$_3$Te$_4$ to induce superconductivity. We could simultaneously identify all three quantum phases, superconductivity, magnetism, and the topological Hall effect in Sn$_{0.06}$Cr$_3$Te$_4$. The pristine Cr$_3$Te$_4$ shows a topological Hall resistivity ($\rho^T_{yz}$) of 240 $n\Omega-cm$, but the Sn intercalation significantly reduces it to 16 $n\Omega-cm$.  In addition, despite the magnetism and Meissner effects being anisotropic, the superconductivity observed from the in-plane electrical resistivity ($\rho_{\it{bb}}$)  is nearly isotropic between $H\parallel \it{bc}$ and $H\parallel \it{a}$, suggesting separate channels of conduction electrons responsible for the superconductivity and magnetism of this system, which is supported by our spin-resolved DFT calculations. We identify two orders of higher carrier density in superconducting Sn$_{0.06}$Cr$_{3}$Te$_4$ than the parent Cr$_3$Te$_4$. A jump in the specific heat is noticed around the $T_c$ with a volume fraction of 33\%, confirming the bulk superconductivity in Sn$_{0.06}$Cr$_{3}$Te$_4$. Importantly, for the first time, this study demonstrates superconductivity in a skyrmion lattice, offering a new class of topological quantum materials.

\begin{figure*}[t]
    \centering
    \includegraphics[width=\linewidth]{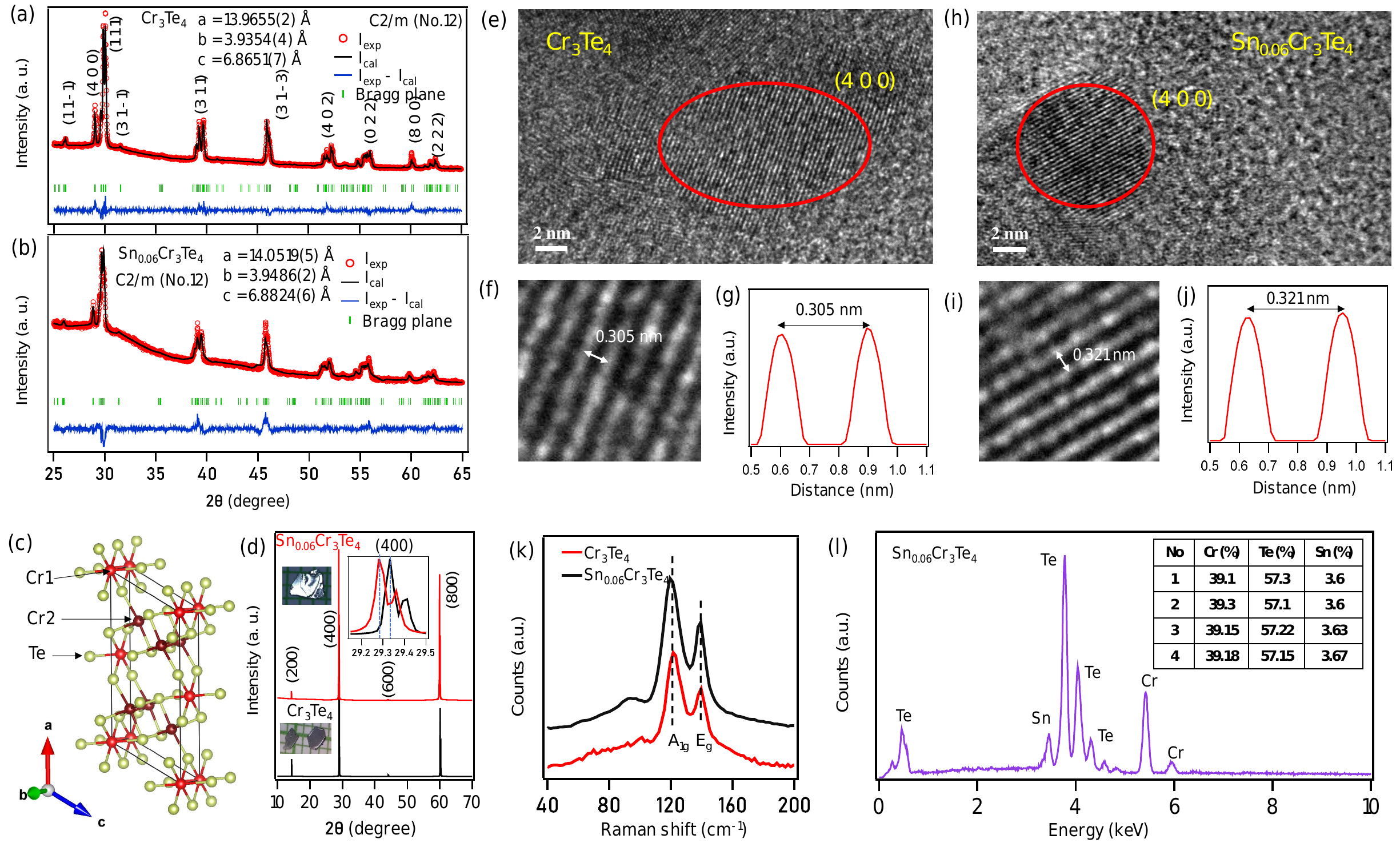}
    \caption{(a) and (b) XRD patterns of the crushed single crystals of Cr$_3$Te$_4$ and Sn$_{0.06}$Cr$_3$Te$_4$, respectively,  overlapped with Rietveld refinement. (c) Schematic image of the monoclinic crystal structure of Cr$_3$Te$_4$. (d)  XRD patterns of Cr$_3$Te$_4$ (bottom panel) and Sn$_{0.06}$Cr$_3$Te$_4$ (top panel) single crystals. Top-middle inset in (d) shows zoomed-in (4 0 0) Bragg's peak demonstrating the lattice shift between the parent and the Sn intercalated samples. Optical images of typical single crystals are shown in the left-side insets of (d). (e) HRTEM image taken on Cr$_3$Te$_4$ crystallite, demonstrating the lattice planes corresponding to the (4 0 0) plane. (f) Zoomed-in HRTEM image from (e), displaying the interplanar distance of 0.305 nm. (g) (4 0 0) lattice plane intensity plot confirming the interplanar distance. (h)-(j) Same as (e)-(g) but from Sn$_{0.06}$Cr$_3$Te$_4$ crystallite. (k) Raman spectra measured on Cr$_3$Te$_4$ and Sn$_{0.06}$Cr$_3$Te$_4$ single crystals. (l) EDS spectra measured on four different Sn$_{0.06}$Cr$_3$Te$_4$ crystallites, confirming the uniform chemical composition. Atomic weight percentages of respective elements are tabulated in the inset of (l).}
    \label{fig1}
\end{figure*}

\begin{figure*}[t]
\includegraphics[width=\linewidth]{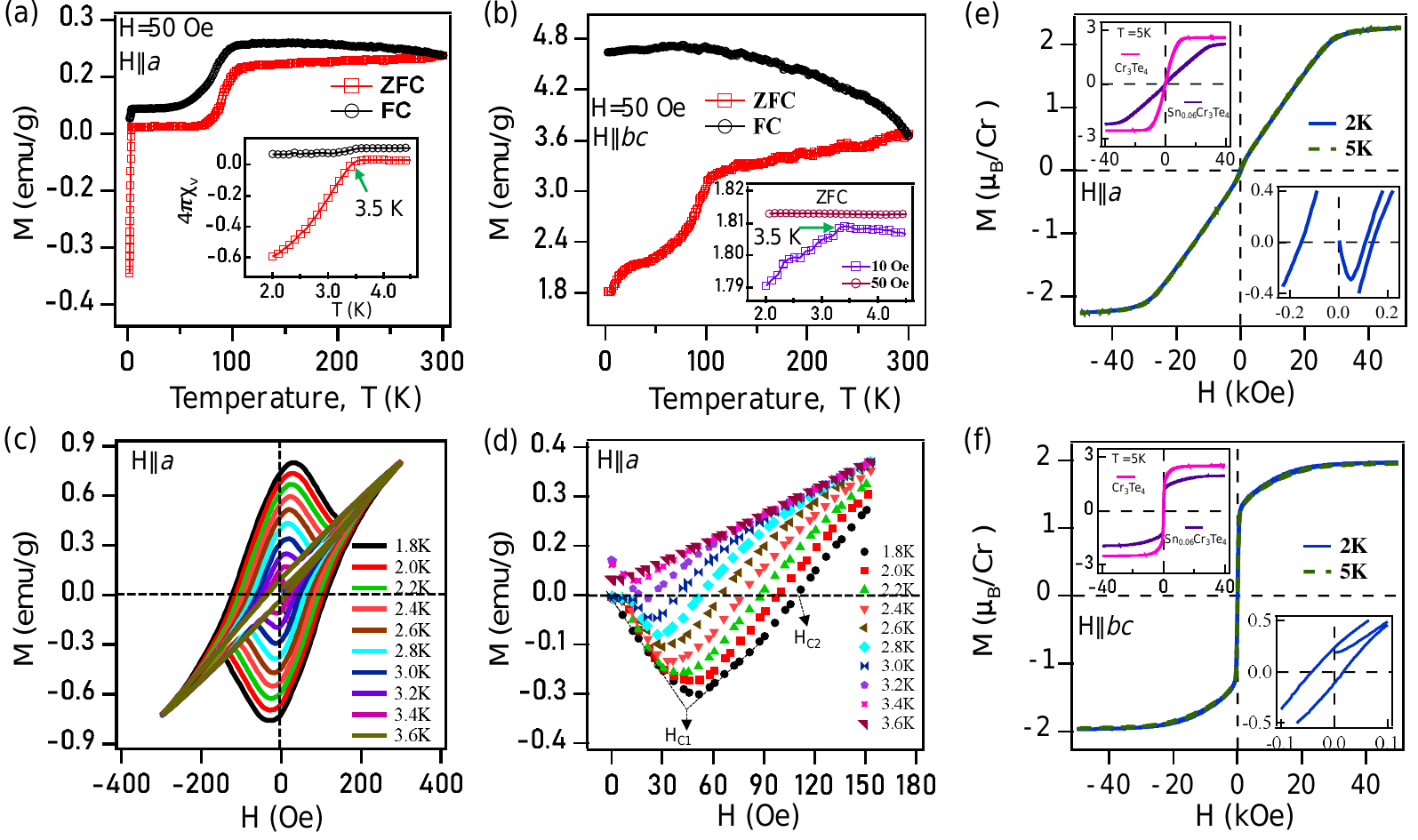}
\caption{Magnetic properties of Sn$_{0.06}$Cr$_3$Te$_4$ single crystal. (a) Temperature dependent magnetization [$M(T)$] measured with H = 50 Oe for $H\parallel \it{a}$. Inset of (a) shows the susceptibility near the superconducting region.  (b) Same as (a) but measured for H$\parallel \it{bc}$. Inset of (b) shows overlapped $M(T)$ measured with 10 and 50 Oe. (c) and (d) Show low field Isothermal magnetization [$M(H)$] curves, demonstrating the Meissner effect within the temperature range of 1.8-3.6 K. $H_{c1}$ and $H_{c2}$ in (d) represent lower and upper critical fields, respectively. (e) $M(H)$ curve measured at 2 and 5 K for $H\parallel \it{a}$. Top-left inset in (e) shows $M(H)$ curves measured at 5 K from Cr$_3$Te$_4$ and Sn$_{0.06}$Cr$_3$Te$_4$ single crystals for $H\parallel \it{a}$. Bottom-right inset of (e) shows zoomed-in $M(H)$, demonstrating the Meissner effect in presence of significant coercive field. (f) $M(H)$ curve measured at 2 and 5 K for $H\parallel \it{bc}$. Top-left inset in (f) shows $M(H)$ curves measured at 5 K from Cr$_3$Te$_4$ and Sn$_{0.06}$Cr$_3$Te$_4$ single crystals for $H\parallel \it{bc}$. Bottom-right inset of (f) shows zoomed-in $M(H)$, demonstrating the absence of Meissner effect for $H\parallel \it{bc}$. }
\label{fig2}
\end{figure*}

\section{Methodology}

\subsection{Experimental Details}
Single crystals of Sn intercalated Cr$_3$Te$_4$ were grown by the Sn-flux method by mixing high purity elements of Cr (5N, Alfa Aesar) and Te (5N, Alfa Aesar) powders as per the stoichiometric ratio in an alumina crucible, added with Sn (99.998\%, Alfa Aesar) shots, and sealed the crucible inside a quartz ampoule under argon atmosphere. A schematic diagram of the single crystal growth heat treatment is shown in Fig. S1 of the supplemental information. Plate-like single crystals with a typical size of 3$\times$2 mm$^2$ with a thickness of 0.15 mm were collected after the reaction. Single crystals of Cr$_3$Te$_4$ were grown by the chemical vapor transport (CVT) technique with iodine as a transport agent as per the procedure described earlier~\cite{Hashimoto1971}. Structural and elemental characterizations were performed using an X-ray diffractometer (Rigaku-9kW, Cu K$_{\alpha}$ of 1.54059 \AA~  wavelength at Room temperature) and energy dispersive spectroscopy (EDS) techniques, respectively.   Temperature-dependent electrical transport, Hall effect, and magnetization measurements were performed using the 9 Tesla physical property measurement system (PPMS, Quantum Design, DynaCool). Electrical resistivity and Hall effect measurements were performed in the standard four-probe method. Temperature-dependent heat capacity measurements were taken using the relaxation method using the Quantum Design-PPMS system. The high-resolution transmission electron microscopy (HRTEM) was done in an FEI, TECNAI TF 20, S-TWIN microscope operated at 200 KV. The samples were prepared by drop-casting onto a carbon-coated copper grid of a 3 mm diameter. Raman spectra were captured by using a micro-Raman spectrometer (LabRam HR Evolution HORIBA France SAS) equipped with a 532 nm laser.

\subsection{DFT Calculations}
Spin-polarized DFT calculations have been carried out on Cr$_3$Te$_4$ and Sn$_{0.06}$Cr$_{3}$Te$_4$ (one unit cell consists of 1 Sn, 6 Cr, and 8 Te atoms) based on the projector augmented wave (PAW) method~\cite{PhysRevB.50.17953} as implemented in the Quantum Espresso package~\cite{giannozzi2009quantum,giannozzi2017advanced}. For the exchange-correlation interaction, we considered the Perdew-Burke-Ernzerhof~\cite{PhysRevLett.77.3865} form of the generalized gradient approximation (GGA). To consider the van der Waals forces, semi-empirical Grimme’s DFT-D2 correction~\cite{grimme2006semiempirical} was considered. To optimize the atomic positions, a cutoff value of 10$^{-3}$ Ry/Bohr was chosen for the Hellmann-Feynmann forces. A plane wave cutoff of 50 Ry was employed. $4\times16\times8$ and $8\times32\times16$ k-mesh with $\Gamma$-centered were used for the self-consistency and DOS calculations, respectively.

\section{Results and Discussion}

$\textit{\textbf{Structural~Properties:-}}$ The exact chemical composition of the Sn intercalated and the parent single crystals were found to be Sn$_{0.06}$Cr$_{2.74}$Te$_4$ and Cr$_{2.76}$Te$_4$, respectively, using the EDS measurements.  For convenience, we use the nominal composition formula of Cr$_3$Te$_4$. Fig.~\ref{fig1}(a) shows the XRD pattern of the crushed Cr$_3$Te$_4$ single crystals measured at room temperature. All peaks in the XRD pattern can be attributed to the monoclinic crystal structure of the $C2/m$ space group (No.12). This is consistent with the crystal structure of Cr$_3$Te$_4$~\cite{Yamaguchi1972, ohta1985magnetic,babot1973proprietes}. Rietveld refinement further confirms the monoclinic structure with lattice parameters,
$\it{a}$=13.9655(2) \AA, $\it{b}$=3.9354(4) \AA, $\it{c}$=6.8651(7) \AA,
$\alpha$=$\beta$=90$^{\circ}$, and $\gamma=118.326(7)^{\circ}$ with $\chi^2$=2.95. Similarly, Fig.~\ref{fig1}(b) shows the XRD pattern of the  crushed Sn$_{0.06}$Cr$_3$Te$_4$ single crystals measured at room temperature. Again, all peaks in the XRD pattern can be attributed to the monoclinic crystal structure of the $C2/m$ space group without any Sn impurity phases.  Rietveld refinement further confirms the monoclinic structure with lattice parameters, $\it{a}$=14.0551(4) \AA, $\it{b}$=3.9494(3) \AA, $\it{c}$=6.8839(5) \AA,
$\alpha$=$\beta$=90$^{\circ}$, and $\gamma=118.376(7)^{\circ}$ with $\chi^2$=3.35. We can see that the Sn intercalation slightly increases the $a$-axis lattice parameter, while the change in lattice parameters of $b$ and $c$ is negligible. Fig.~\ref{fig1}(c) depicts the schematic crystal structure of Cr$_3$Te$_4$ in which the alternative staking of Cr(1) and Cr(2) layers is demonstrated along the $a$-axis.

\begin{figure*}[t]
\includegraphics[width=\linewidth]{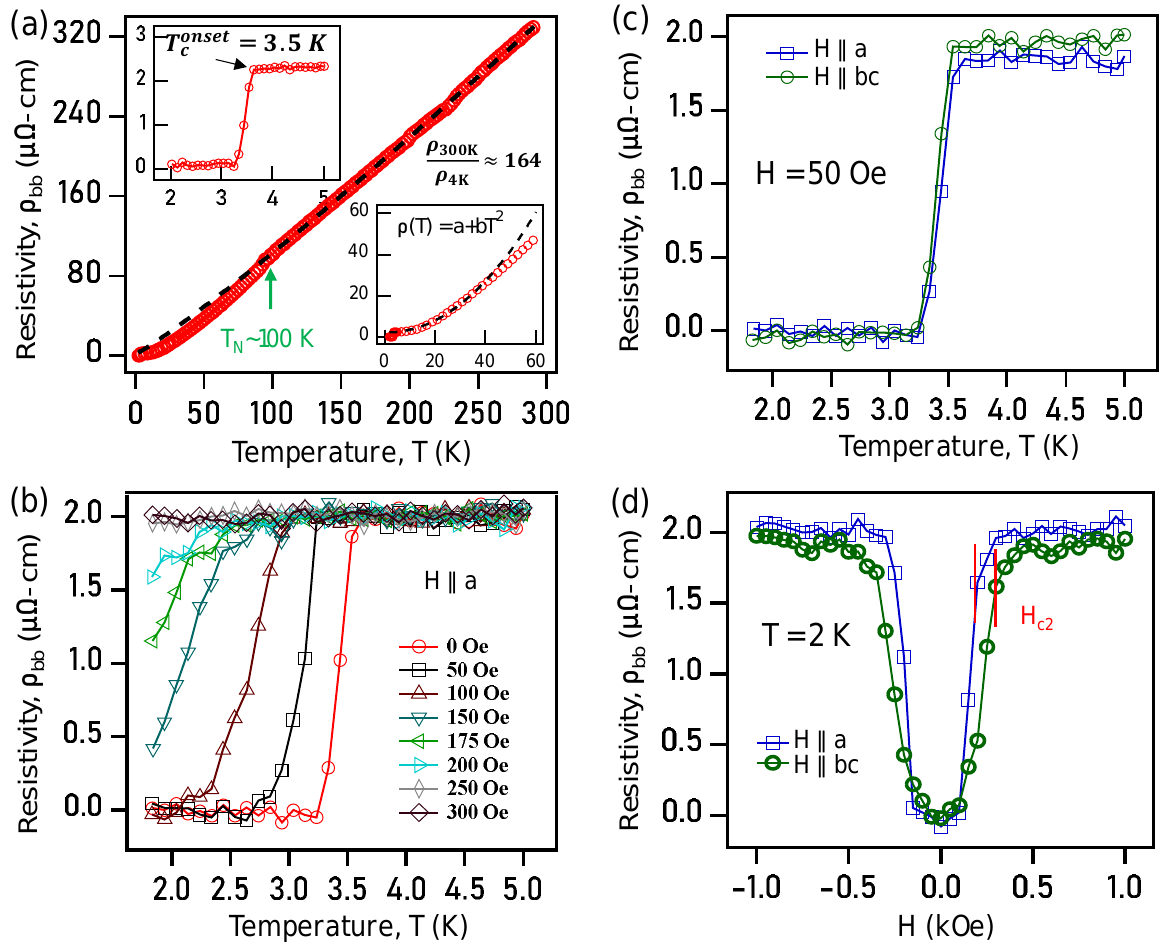}
\caption{Electrical transport properties of Sn$_{0.06}$Cr$_3$Te$_4$ single crystal. (a) Temperature dependent in-plane electrical resistivity ($\rho_{\it{bb}}(T)$). Top inset of (a) shows onset superconducting transition of $T_c\approx3.5$ K. Bottom inset of (a) represents Fermi-liquid fitting at low temperature region. (b) $\rho_{\it{bb}}(T)$ measured under various magnetic field within temperature range of 1.8-5 K for $H\parallel \it{a}$. (c) $\rho_{\it{bb}}(T)$ measured within temperature range of 1.8-5 K for $H\parallel \it{a}$ and $H\parallel \it{bc}$ at an applied field of 50 Oe. (d) $\rho_{\it{bb}}(H)$ measured as a function of field for $H\parallel \it{a}$ and $H\parallel \it{bc}$ at 2 K.}
\label{fig3}
\end{figure*}

The XRD patterns taken on the single crystal of  Cr$_3$Te$_4$ and Sn$_{0.06}$Cr$_{3}$Te$_4$ are shown in the bottom and top panels of Fig.~\ref{fig1}(d), respectively, which are consistent with the monoclinic space group of $C2/m$, suggesting that the crystal growth axis is parallel to the $\it{a}$-axis~\cite{ohta1985magnetic, babot1973proprietes, Purwar2023}. Inset in Fig~\ref{fig1}(d) shows the overlapped $(4~0~0)$ reflections from Cr$_3$Te$_4$ and Sn$_{0.06}$Cr$_{3}$Te$_4$ in which the peak position is shifted towards the lower 2$\theta$ for Sn$_{0.06}$Cr$_3$Te$_4$ compared to Cr$_3$Te$_4$. The XRD peak shifting to a lower 2$\theta$ value indicates an expansion of the interplanar spacing along the $\it{a}$-axis due to Sn intercalation~\cite{adam2021superconducting,naik2019effect}. Fig.~\ref{fig1}(e) shows the HRTEM image taken on Cr$_3$Te$_4$ crystallite, demonstrating the lattice planes corresponding to the $(4~0~0)$  plane. Fig.~\ref{fig1}(f) depicts the zoomed-in HRTEM image of Fig.~\ref{fig1}(e), displaying the interplanar distance of 0.305 nm which is in agreement with the interplanar spacing of 0.307 nm [$(4~0~0)$] derived from the Rietveld refinement. This is further confirmed from the lattice plane intensity plot as shown in Fig.~\ref{fig1}(g). Figs.~\ref{fig1}(h)-(j) depicts information of $(4~0~0)$ plane as similar as shown in Figs.~\ref{fig1}(e)-(g) but from a Sn$_{0.06}$Cr$_3$Te$_4$ crystallite. From Figs.~\ref{fig1}(i) and ~\ref{fig1}(j), we can clearly notice that Sn intercalation increases the $(4~0~0)$ interplanar spacing of about 0.016 nm compared to the parent system. This is in good agrement with the XRD peak shifting to lower 2$\theta$ angle with Sn intercalation [see inset in Fig.~\ref{fig1}(d)].

Next, from the Raman spectra as shown in Fig.~\ref{fig1}(k), we mainly observe two prominent phonon peaks, positioned at approximately 123.5 cm$^{-1}$ and 139.79 cm$^{-1}$. These peaks correspond to the distinct vibrational modes such as the out-of-plane A$_{1g}$ and the in-plane E$_g$, of a typical Cr$_x$Te$_y$ system~\cite{Chen2022,Li2022}. This observation confirms that the intercalated Sn atoms take the positions in the Cr(2)-intercalated layer [see Fig.~\ref{fig1}(c)].  Importantly, we do not observe any additional Raman modes corresponding to the Sn impurity which shall be at around 202.5 cm$^{-1}$ ~\cite{Wang1997}. Further, to be more accurate on the chemical composition, particularly, of the Sn intercalated system, we performed EDS measurements on several crystallites of Sn$_{0.06}$Cr$_3$Te$_4$  as shown  Fig.~\ref{fig1}(l). From Fig.~\ref{fig1}(l), it is evident that the chemical composition is uniform across all the measured crystallites within the error-bars (Sn$_{0.063(2)}$Cr$_{2.74(1)}$Te$_4$).

$\textit{\textbf{Magnetic~Properties:-}}$ Magnetization [$M(T)$], plotted as a function of temperature for Sn$_{0.06}$Cr$_{3}$Te$_4$  both field-cooled (FC) and zero-field-cooled (ZFC) modes, measured with a magnetic field of 50 Oe applied parallel to $\it{a}$-axis ($H\parallel \it{a}$) is shown in Fig.~\ref{fig2}(a) and the field applied parallel to $\it{bc}$-plane ($H\parallel \it{bc}$) is shown in Fig.~\ref{fig2}(b). The Meissner effect is clearly visible from the ZFC data of $H\parallel \it{a}$, as shown in the inset of Fig.~\ref{fig2}(a), at an onset transition temperature of 3.5 K, below which the magnetization becomes negative due to diamagnetism in the superconducting phase. In addition to the Meissner effect at 3.5 K, we also identify a sharp decrease in magnetization from FC and ZFC modes at around 100 K, possibly due to a canted antiferromagnetic (AFM) transition~\cite{andresen1970magnetic}. In contrast to $H\parallel \it{a}$, the magnetization measured for $H\parallel \it{bc}$ shows a different character as demonstrated in Fig.~\ref{fig2}(b). Firstly, we do not observe the Meissner effect below 3.5 K from the ZFC data, and secondly, only the ZFC data shows a AFM transition at around 100 K, but not the FC data. In fact, the saturated magnetization derived from the FC data ($M_S\approx$4.65 emu/g) at 2 K for $H\parallel \it{bc}$ is almost 20 times higher than the saturation magnetization ($M_S\approx$0.23 emu/g) of  $H\parallel \it{a}$, indicating a large magnetic anisotropy present in the system~\cite{jiao2011anisotropic}. This, further implies that the $\it{bc}$-plane is the easy magnetization plane with stronger magnetic exchange interactions, suppressing the superconducting state in the $\it{bc}$ plane. In support of this, the $M(T)$ data measured with an applied field of 10 Oe for $H\parallel \it{bc}$ [see the inset of Fig.~\ref{fig2}(b)] shows a slight decrease in magnetization precisely at the onset temperature of 3.5 K, suggesting the coexistence of both magnetism and superconductivity in the $\it{bc}$-plane though the stronger ferromagnetic exchange interactions dominate the Meissner effect.

\begin{figure}[t]
\centering
\includegraphics[width=\linewidth]{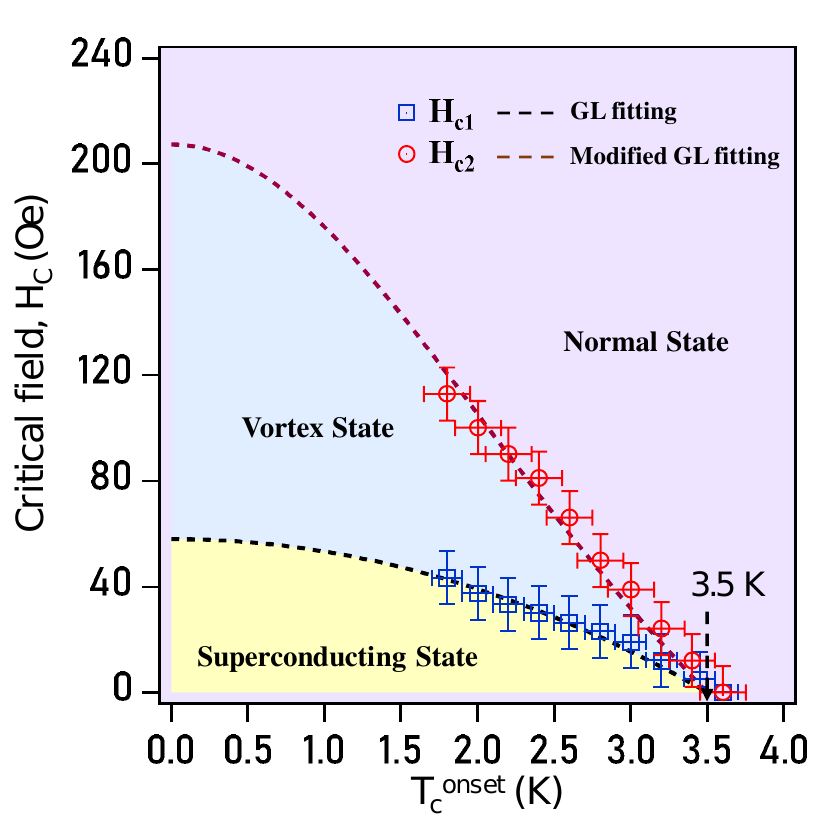}
\caption{Temperature dependence of lower critical field $H_{c1}$ and upper critical field $H_{c2}$ along with schematic diagram of shaded superconducting, vortex, and normal state regions.}
\label{fig4}
\end{figure}

 Isothermal magnetization curves [$M(H$)] across the superconducting transition temperature for $H\parallel \it{a}$ are shown in Fig.~\ref{fig2}(c). Fig.~\ref{fig2}(d) shows the zoomed-in data of Fig.~\ref{fig2}(c) taken for positive applied fields. From Fig.~\ref{fig2}(d),  we notice that at lower fields, the negative magnetization increases linearly with the field up to a lower critical field of $H_{c1}$. However, beyond a certain field,  the negative magnetization starts decreasing with increasing field and approaches zero magnetization at an upper critical field of $H_{c2}$. This observation is consistent at all measured temperatures up to 3.4 K, indicating Sn$_{0.06}$Cr$_3$Te$_4$ to be a type-II superconductor~\cite{Tinkham2017}. Interestingly, at 3.4 K, despite the magnetization initially decreasing with increasing field like in a typical superconductor, unusually, it is always positive. This could be due to the coexistence of magnetism and superconductivity~\cite{Ishida2021}.   Figs.~\ref{fig2}(e) and ~\ref{fig2}(f) show the magnetization isotherms measured at 2 and 5 K for both $H\parallel \it{a}$ and $H\parallel \it{bc}$, respectively. These data show that the out-of-plane magnetization ($\it{a}$-axis) linearly increases with the field up to 3 T and then saturates to 2.2 $\mu_B$/Cr. The bottom-right inset of Fig.~\ref{fig2}(e) measured at 2 K again demonstrates the superconducting state at lower fields together with a significant coercivity of 150 Oe.

 The top-left inset of Fig.~\ref{fig2}(e) shows the magnetization isotherms [$M(H)$] measured at 5 K from both parent and the superconducting  Sn$_{0.06}$Cr$_3$Te$_4$ for $H\parallel \it{a}$. From this data, we can see that the magnetization of Cr$_3$Te$_4$ saturates faster than that of Sn$_{0.06}$Cr$_3$Te$_4$.    On the other, the in-plane magnetization ($\it{bc}$-plane) of Sn$_{0.06}$Cr$_3$Te$_4$ spontaneously saturates to 2 $\mu_B$/Cr under the applied magnetic field as shown in  Fig.~\ref{fig2}(f), indicating that the $\it{bc}$-plane is the easy-plane of magnetization similar to the parent system~\cite{HAMASAKI1975895, Dijkstra_1989}.  The bottom-right inset of Fig.~\ref{fig2}(f) shows the $M(H)$ data measured at 2 K,  demonstrating an absence of a superconducting state in the $bc$-plane even at very low applied fields. The top-left inset of Fig.~\ref{fig2}(f) shows magnetization isotherms [$M(H)$] measured at 5 K  from both Cr$_3$Te$_4$ and the superconducting  Sn$_{0.06}$Cr$_3$Te$_4$ for $H\parallel \it{bc}$. This data shows that the Sn intercalation has no effect on the in-plane magnetization of Cr$_3$Te$_4$ except for a slight reduction in the saturated magnetic moment.

\begin{figure*}[t]
\includegraphics[width=\linewidth]{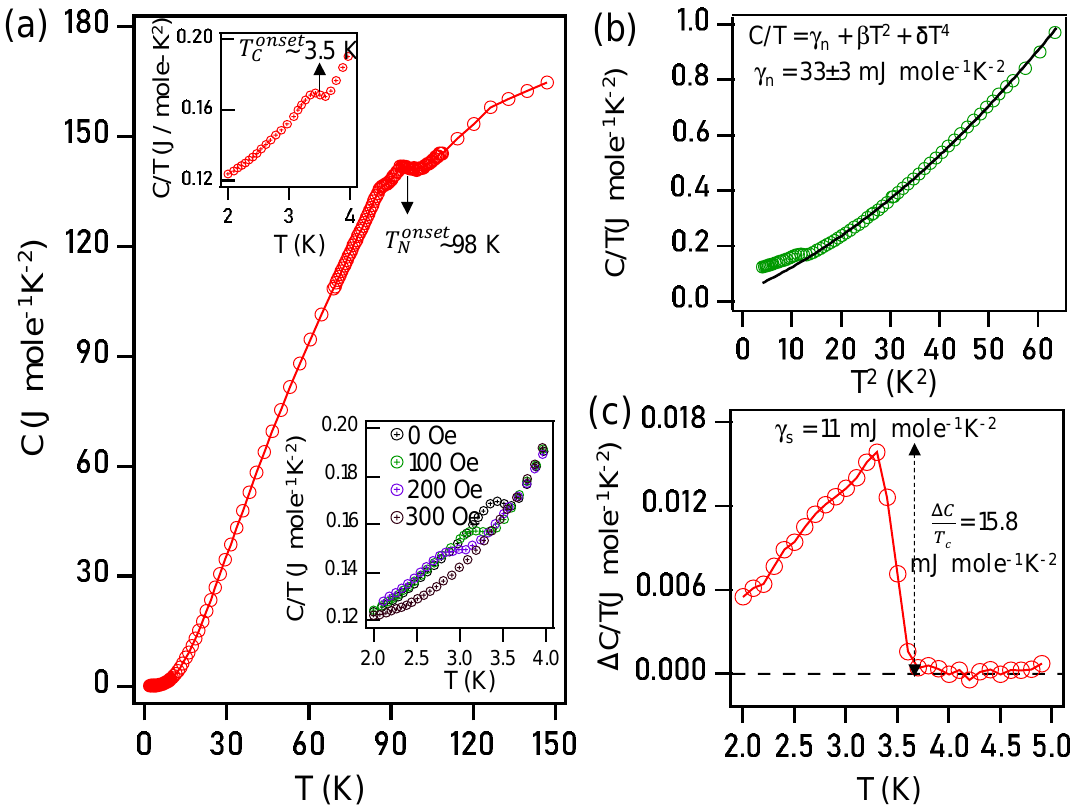}
\caption{ Specific heat $C(T)$ measurements on Sn$_{0.06}$Cr$_3$Te$_4$ single crystal. (a) Temperature dependent $C(T)$. Top-right inset of (a) shows zoomed-in $C(T)/T$ taken around the $T_c$.  Bottom-left inset of (a) shows $C(T)/T)$ measured under magnetic applied fields.  (b) $C/T$ data fitted using the relation $C(T)/T=\gamma_{n}+\beta T^2 + \delta T^4$ to extract the Sommerfeld coefficient in the normal state. (c) Depicts ${\Delta C}/{T}$ plotted as a function of temperature.}
\label{fig3a}
\end{figure*}

$\textit{\textbf{Electrical~Properties:-}}$ In-plane ($\rho_{bb}$) electrical resistivity of Sn$_{0.06}$Cr$_{3}$Te$_4$ plotted as a function of temperature is shown in Fig.~\ref{fig3}(a). At around 100 K, we identify a hump in the resistivity curve due to the AFM transition. The high-temperature resistivity between 100 and 300 K is linearly dependent on the temperature, as the black-dashed-line fitting demonstrates. The top-left inset of Fig.~\ref{fig3}(a) demonstrates a drop in resistivity to zero at an onset superconducting transition temperature of $T_c\approx3.5$ K. The bottom-right inset of Fig.~\ref{fig3}(a) suggests Fermi-liquid type electrical resistivity at very low temperatures ($<$ 40 K), as the resistivity quadratically depends on the temperature. In addition, the large residual resistivity ratio (RRR) of $\frac{\rho_{300K}}{\rho_{4K}}$$\approx 164$ confirms the high quality of the studied samples. Fig.~\ref{fig3}(b) shows the electrical resistivity measured by varying the field strength. In Fig.~\ref{fig3}(b), a gradual decrease in the onset $T_c$ is noticed with increasing the field. Interestingly, despite the anisotropic magnetization, the superconducting transition temperature is found to be unchanged between $H\parallel \it{a}$ and $H\parallel \it{bc}$ as $\rho_{\it {bb}}(T)$ shows equal onset $T_c$ of 3.5 K when measured with 50 Oe [see Fig.~\ref{fig3}(c)]. Fig.~\ref{fig3}(d) depicts $\rho_{\it {bb}}$ plotted as a function magnetic field for $H\parallel \it{a}$ and $H\parallel \it{bc}$  at 2 K.  We observe that the upper critical field of $H_{c2}$$\approx$300 Oe is higher for $H\parallel \it{bc}$ compared to $H_{c2}$$\approx$200 Oe for $H\parallel \it{a}$.

\begin{figure*}[t]
    \centering
    \includegraphics[width=\linewidth]{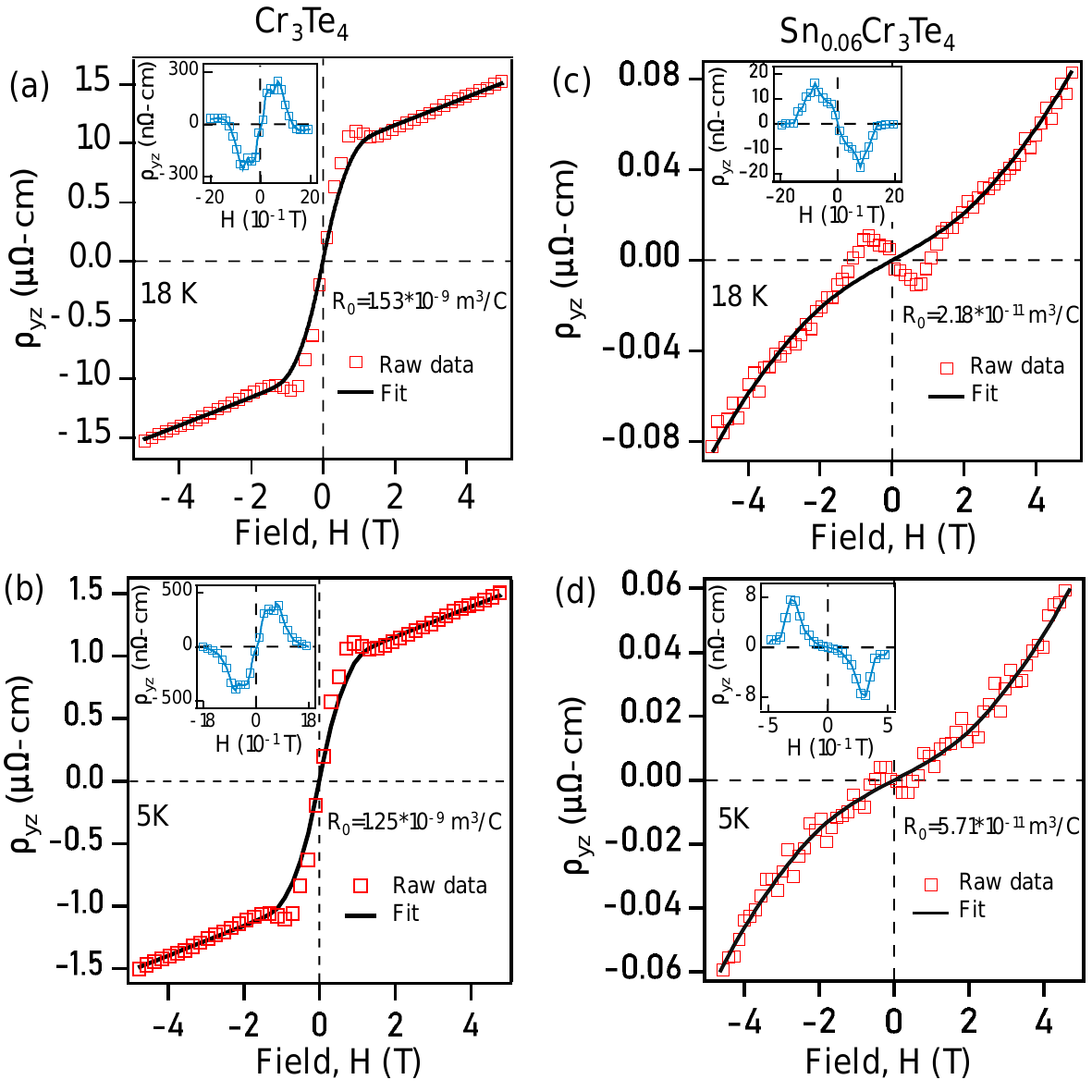}
    \caption{(a) and (b) Hall resistivity ($\rho_{yz}$) plotted as a function of field for Cr$_3$Te$_4$ measured at 1.8 and 5 K, respectively. (c) and (d) are same as (a) and (b) but measured on Sn$_{0.06}$Cr$_{3}$Te$_4$ (b). The black-curves are fits to the data. Insets in (a)-(d) show the topological Hall resistivity plotted as a function of field (see the text for more details).}
    \label{fig6}
\end{figure*}

\begin{figure*}[t]
    \centering
     \includegraphics[width=\linewidth]{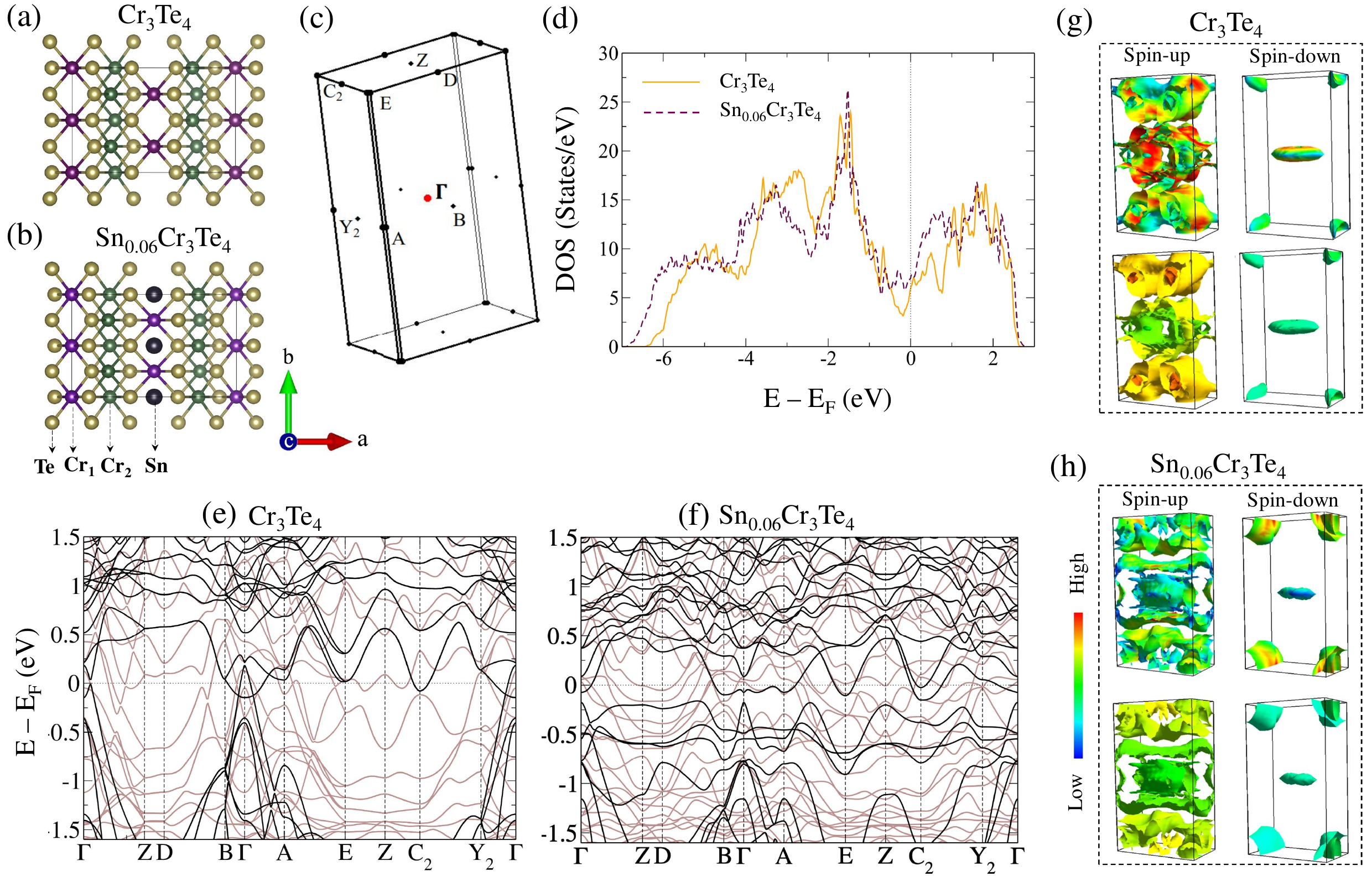}
    \caption{Crystal structure of parent Cr$_3$Te$_4$  (a) and Sn$_{0.06}$Cr$_3$Te$_4$ (b) projected onto the $ab$-plane. Magenta, green, yellow and black spheres represent Cr1, Cr2, Te, and Sn atoms, respectively.  (c) High symmetry points defined on the monoclinic Brillouin zone. (d) Yellow solid and maroon dashed lines denote the density of states (DOS) for the parent Cr$_3$Te$_4$ and Sn$_{0.06}$Cr$_3$Te$_4$ systems, respectively. Spin polarised band structures for the parent Cr$_3$Te$_4$ (e)  and Sn intercalated Cr$_3$Te$_4$ (f).  Brown and black colours in (e) and (f) denote the bands for the up and down spins, respectively. (g) Fermi surfaces of Cr$_3$Te$_4$  plotted for the spin-up and spin-down channels. (h) Fermi surfaces of Sn intercalated Cr$_3$Te$_4$  plotted for the spin-up and spin-down channels. The color scale in (g) and (h) represent the Fermi velocity.}
    \label{fig5}
\end{figure*}

Critical fields ($H_{c1}$ and $H_{c2}$) and critical temperatures ($T_c$) estimated from Fig.~\ref{fig2}(d) are plotted in Fig.~\ref{fig4}. The data of $H_{c1}(T)$ is reasonably fitted using the Ginzburg-Landau equation of quadratic field dependence on the temperature~\cite{ginzburg1957antagonie},  $H_{\mathrm{c}1}(T)=H_{\mathrm{c}1}(0)[1-(T/T_{\mathrm{c}})^2]$.   From the fitting, we obtained a lower critical field of $H_{\mathrm{c}1}(0)$ = $58\pm 4$ Oe at zero temperature. On the other hand, the upper critical field of $H_{c2}(T)$ is reasonably fitted using the modified Ginzburg-Landau model which takes the form $H_{\mathrm{c}2}(T)=H_{\mathrm{c}2}(0)[(1-t^2)/(1+t^2)]$, where $t=T/T_{\mathrm{c}}$. From this fitting, we obtained an upper critical field of  $H_{\mathrm{c}2}(0)$ = $205\pm 10$ Oe at zero temperature. The Ginzburg-Landau coherence length,  $\xi(0)=1264.8\pm90$ \AA,  is estimated using the relation $H_{\mathrm{c}2}(0)=\phi_0/2\pi{\xi(0)}^2$, where $\phi_0$ is the flux quanta. The mean critical field H$_c$ is quantified to $109\pm 9$ Oe using the relation, $\mathrm{H_c}=\sqrt{\mathrm{H_{c1}H_{c2}}}$. Now, by knowing the upper critical field $H_{\mathrm{c}2}(0)$ and mean critical field $\mathrm{H_c}$ one can estimate the GL parameter ($\kappa$) using the relation, $\mathrm{H_{c2}}(0)=\sqrt{2}\kappa\mathrm{H_{c}}$. Thus, the obtained $\kappa=1.33$ in this study is well above the threshold value of $\frac{1}{\sqrt{2}}$, again confirming the type-II superconductivity of this system. With the help of GL coherence length ($\xi(0)$) and GL parameter ($\kappa$), we further estimated the London penetration depth ($\lambda_{L}$) equal to $1682.1\pm120$ \AA  ~at zero temperature,  using the relation $\lambda_{L}(0)=\kappa\xi(0)$. This value is much larger than any other ferromagnetic superconductor. The mean-free path ($\it{l}$) of the charge carriers is estimated using the relation $\it{l}=h{k_F}/2\pi\rho_0ne^2$, where $k_F$ is approximated to $(3{\pi}^2n)^{1/3}$ by considering a spherical Fermi surface. Inserting the values, $\rho_0=1.78\times 10^{-8}\: \Omega$-m and $n=1.09\times 10^{29}\: m^{-3}$, we obtained $\it{l}=313.4$ \AA. Thus, as $\it{l}<< \xi(0)$, Sn$_{0.06}$Cr$_{3}$Te$_4$ falls in to the dirty-limit of superconductivity~\cite{Parks2018}.

$\textit{\textbf{Specific Heat Capacity:-}}$ To confirm the bulk superconductivity of Sn$_{0.06}$Cr$_{3}$Te$_4$, we performed specific heat [$C(T)$] measurements within the temperature range of 2-150 K as shown in Fig.~\ref{fig3a}(a). A specific heat jump at an onset transition temperature $T_c\approx3.5$ K is noticed from the top-left inset of Fig.~\ref{fig3a}(a), which is in agreement with the magnetic and transport data shown in Fig.~\ref{fig2}(a) and Fig.~\ref{fig3}, respectively. The bottom-right inset of Fig.~\ref{fig3a}(a) depicts the specific heat data measured under various magnetic fields for $H\parallel a$. The $T_c$ and the height of the specific heat jump decrease with increasing the field, and the superconductivity is completely suppressed at an applied field of 300 Oe. In addition, we also observe a specific heat jump around the magnetic transition of 98 K. In Fig.~S3(b) of the supplemental information, we provided field-dependent $C(T, H)$ data from which the magnetic transition at 98 K is concluded to be an AFM-type. From Fig.~\ref{fig3a}(a), we observe the Debye-type increase in heat capacity with temperature up to 150 K, and beyond 150 K it reaches the Dulong-Petit limit. In agreement to this observation, the calculated heat capacity $C=176.1$ J mole$^{-1}$ K$^{-1}$, following the Dulong-Petit law [$C=3nR$], is close to the experimental heat capacity value at 150 K ($C=165 $ J mole$^{-1}$ K$^{-1}$). Here $n=7.06$ is the number of atoms per formula unit (Sn$_{0.06}$Cr$_3$Te$_4$) and $R=8.314$ J mole$^{-1}$ K$^{-1}$ is the universal gas constant.

Further, we could reasonably fit the normal state heat capacity data, taken at zero fields, following the relation $C(T)=\gamma_{n}T+\beta T^3+\delta T^5$ [see Fig.~\ref{fig3a}(b)]. Here, the first term ($\gamma_{n}T$) represents the electronic contribution, and the second and third terms ($\beta T^3+\delta T^5$) represent the phonon contribution to the total heat capacity. From the fitting, we derived a Sommerfeld coefficient ($\gamma_{n}$) of the electronic specific heat in the normal state $\gamma_{n}=33\pm3$ mJ mole$^{-1}$ K$^{-2}$ and the coefficients related to the phonon contribution (Debye constants) $\beta=8\pm$0.2 mJ mole$^{-1}$ K$^{-4}$ and $\delta=0.1\pm$0.04 mJ mole$^{-1}$ K$^{-6}$. We also estimated the Sommerfeld coefficient in the superconducting state $\gamma_{s}=11$ mJ mole$^{-1}$ K$^{-2}$ [see Fig.~\ref{fig3a}(c)] using the BCS equation of the weak-coupling limit, ${\Delta C}/{T_c}=[C(H=0~ Oe)-C(H=300~ Oe)]/{T_c} = 1.43*\gamma_{s}$. Finally, the superconducting volume fraction is estimated using the relation $\frac{\gamma_{s}}{\gamma_{n}}\times100\approx33\pm4\%$~\cite{Kim2006}, which is much larger than the Sn concentration ($6\%$) present in Sn$_{0.06}$Cr$_{3}$Te$_4$. Thus, the large SC volume fraction rules out the Sn impurity phase superconductivity.

$\textit{\textbf{Topological Hall Effect:-}}$  The parent Cr$_3$Te$_4$ is known to show significant topological Hall effect (THE) originated by the skyrmion lattice~\cite{Liu2017, Zhu2018, Purwar2023}.   In order to uncover the presence of THE in the superconducting Sn$_{0.06}$Cr$_{3}$Te$_4$, we performed Hall measurements on both Cr$_3$Te$_4$ and Sn$_{0.06}$Cr$_{3}$Te$_4$ as shown in Fig.~\ref{fig6}. Figs.~\ref{fig6}(a) and ~\ref{fig6}(b) show the field dependent total Hall resistivity $\rho_{yz}(H)$ of Cr$_3$Te$_4$ measured at 1.8 K and 5 K, respectively.  Similarly, Figs.~\ref{fig6}(c) and ~\ref{fig6}(d) show $\rho_{yz}(H)$ of Sn$_{0.06}$Cr$_{3}$Te$_4$ measured at 1.8 K and 5 K, respectively. $\rho_{\it{yz}}$  is measured with current along the $\it{y}$-axis and magnetic field applied along the $\it{x}$-axis to get the Hall voltage along the $\it{z}$-axis.  The total Hall resistivity ($\rho_{\it{yz}}$) may have contributions from the normal Hall effect ($\rho^N$) and the anomalous Hall effect ($\rho^A$). Thus, the total Hall resistivity can be expressed by the empirical formula, $\rho_{\it{yz}}(H)=\rho^N(H)+\rho^A(H)=\mu_0R_0H+\mu_0R_SM$, where $R_0$ and $R_S$ are the normal and anomalous Hall coefficients, respectively. The fitting (black curves) should be nearly perfect if there is no topological Hall contribution. However, from Figs.~\ref{fig6} (a)-(d), we can notice that the fitting is not perfect. Therefore, the topological Hall resistivity also contributes to the total Hall resistivity, which is expressed by $\rho_{\it{yz}}(H)=\rho^N(H)+\rho^A(H)+\rho^T(H)$. Thus, the topological Hall contribution is extracted using the relation, $\rho^T(H)= \rho_{\it{yz}}(H)-[\rho^N(H)+\rho^A(H)]$ ~\cite{Nakatsuji2015, Kuroda2017, Kanazawa2011} which are shown in the insets of Figs.~\ref{fig6} (a)-(d).

Further, as can be seen from the insets of Figs.~\ref{fig6} (a) and (b), at 2 K, a maximum topological Hall resistivity of $\rho^T_{yz}\approx$ 240 $n\Omega-cm$ is found in Cr$_3$Te$_4$ at a critical field of 0.7 T, while  $\rho^T_{yz}\approx$ 16 $n\Omega-cm$ is observed from Sn$_{0.06}$Cr$_{3}$Te$_4$ around at the same critical field of 0.8 T. On the other hand, at 5 K, the maximum topological Hall resistivity of 8 $n\Omega-cm$ is observed in Sn$_{0.06}$Cr$_{3}$Te$_4$ at a critical field of 0.3 T whereas in Cr$_3$Te$_4$ we observe a slight increase in $\rho^T_{yz}$ to 345 $n\Omega-cm$.   An inverted topological Hall signal between Cr$_3$Te$_4$ and Sn$_{0.06}$Cr$_{3}$Te$_4$ [see the insets of Figs.~\ref{fig6}(a) and ~\ref{fig6}(c)] hints at the helicity switching of the skyrmion lattice~\cite{Shibata2013, Yao2020}. These observations demonstrates a tuning of topological properties in Cr$_3$Te$_4$ by the Sn intercalation. Importantly, we successfully demonstrate the topological Hall effect in the presence of superconducting state, implying the coexistence of exotic quantum phases skyrmion lattice and superconductivity. Next, the carrier concentration ($n$) of Cr$_3$Te$_4$ and Sn$_{0.06}$Cr$_{3}$Te$_4$ is calculated by using the formula $n=1/(R_0 |q|)$, where $q$ is the carrier charge. We obtained the carrier concentration ($n$) of 4.07$\times10^{21}$ cm$^{-3}$ and 2.86$\times10^{23}$ cm$^{-3}$ from Cr$_3$Te$_4$ and Sn$_{0.06}$Cr$_{3}$Te$_4$ at 2 K, respectively. On the other hand, at 5 K, these values are given by 5.01$\times10^{21}$ cm$^{-3}$ and 1.10$\times10^{23}$ cm$^{-3}$ from Cr$_3$Te$_4$ and Sn$_{0.06}$Cr$_{3}$Te$_4$, respectively. This clearly indicates that the Sn intercalation enhances the carrier density in Cr$_3$Te$_4$ by an order of two. The hole carrier density of Cr$_3$Te$_4$ estimated in this study is consistent with previous reports~\cite{Topological_Hall_cr2te3,Molecular_beam}.

Several mechanisms are proposed to understand the topological Hall effect in quantum materials. Such as the Dzyaloshinskii–Moriya (DM) interaction in the noncentrosymmetric systems~\cite{yu2010real,jiang2017skyrmions,PhysRevLett.113.087203} or the uniaxial magnetocrystalline anisotropy (MCA) in the centrosymmetric systems~\cite{Rout2019,ding2019observation,gilbert2015realization,doi:10.7566/JPSJ.87.074704}. In our present case of the centrosymmetric Cr$_{3}$Te$_4$, the chiral-spin structure is stabilized by the strong MCA as already reported earlier by the same authors~\cite{Purwar2023}. The chiral-spin structure is crucial to obtain the THE as the itinerant electrons acquire a real-space Berry curvature associated with finite scalar-spin chirality  $\chi_{ijk}=S_{i}.(S_{j}\times S_{k})$ which serves as fictitious magnetic field to generate the topological Hall signal~\cite{PhysRevLett.98.057203,taguchi2001spin,wang2022topological}. An array of chiral-spin structures can form a lattice structure called the skyrmion lattice. Our findings of topological Hall effect in Cr$_3$Te$_4$ and Sn$_{0.06}$Cr$_{3}$Te$_4$ are consistent with earlier studies on  Cr$_x$Te$_y$-type systems demonstrating the presence of skyrmion lattice~\cite{Saha2022, Liu2022, Feng2023, Zhang2023}.

$\textit{\textbf{Electronic Band Structure:-}}$ Our experimental results of increased charge carrier density with Sn intercalation are qualitatively supported by the density functional theory (DFT) calculations. For the DFT calculations, we considered the conventional unit cell of Cr$_3$Te$_4$, consisting of 6 Cr atoms (2 Cr1-type and 4 Cr2-type) and 8 Te atoms per unit cell as shown in Fig.~\ref{fig5}(a). For the Sn intercalated Cr$_3$Te$_4$, we considered one Sn atom per primitive unit cell [see Fig.~\ref{fig5}(b)]. From the DFT calculations, we find that Cr$_3$Te$_4$ is a ferromagnetic metal with an average magnetic moment of 3.32 $\mu_{B}$/Cr. With the intercalation of Sn, the metallicity and ferromagnetic nature persist, but the average magnetic moment slightly reduces to 3.29$\mu_{B}$/Cr. These values are slightly higher than the experimental values of 2.56 $\mu_{B}$/Cr for Cr$_3$Te$_4$~\cite{Yamaguchi1972, Wang2022} and  2.10 $\mu_{B}$/Cr for Sn$_{0.06}$Cr$_{3}$Te$_4$ when measured at 2 K.  Fig.~\ref{fig5}(d) shows the total density of states (DOS) plotted for Cr$_3$Te$_4$ and 6\% Sn intercalated Cr$_3$Te$_4$. We observe that the density of states (DOS) of Sn$_{0.06}$Cr$_3$Te$_4$,   near the Fermi level,  are significantly increased compared to the DOS of Cr$_3$Te$_4$. This behaviour is consistent with the experimental observation of enhanced charge carrier density with Sn intercalation. Moreover, the calculated Fermi energy of $E_F = 9.4650$ eV for Cr$_3$Te$_4$ increases to 10.9425 eV by the Sn intercalation. That means, the Fermi energy is shifted by about 1.5 eV towards the higher kinetic energy due to a significant change in the low-energy electronic structure of Cr$_3$Te$_4$ with the Sn intercalation. We also performed DFT+U calculations~\cite{PhysRevB.52.R5467} for different values of $U$ ranging up to $U = 3$ eV [see Fig. S4 in the supplemental information]. Interestingly, for all the values of $U$ up to 1 eV, the calculations show enhanced DOS near the Fermi energy in Sn$_{0.06}$Cr$_3$Te$_4$ compared to Cr$_3$Te$_4$ [see Fig. S5 in the supplemental information]. However, for $U = 2$ eV and higher we do not find a significant difference in the DOS near $E_F$ between  Sn$_{0.06}$Cr$_3$Te$_4$ and Cr$_3$Te$_4$.

From the spin-resolved PDOS [see Fig. S6 of the supplemental information], we identify that the Cr $3d$ orbitals dominate the spin-up DOS near the Fermi level, while they have a negligible contribution to the spin-down DOS. Specifically, we notice that the spin-up DOS is dominated by $d_{xy}$, $d_{x^2-y^2}$, and $d_{zx}$ orbitals in the vicinity of Fermi level. However, by the Sn intercalation,   we notice a decrease in the spin-up DOS  and drastic increase (about two times) in the spin-down DOS of $d_{x^2-y^2}$ and $d_{zx}$. On the other hand, the change in $d_{xy}$ DOS is almost negligible by the Sn intercalation. This clearly indicates that the orbitals,  $d_{x^2-y^2}$ and $d_{zx}$ are critical for the superconductivity in  Sn$_{0.06}$Cr$_{3}$Te$_4$. Having understood the orbital contribution near the Fermi level, we then went on to examine the band dispersions [see Figs.~\ref{fig5}(e)  and ~\ref{fig5}(f)] and Fermi surfaces [see Figs.~\ref{fig5}(g) and ~\ref{fig5}(h)]. It is evident from the band dispersions that the Sn intercalation not only shifts the Fermi level to higher kinetic energies but significantly alters the band structure near the Fermi level. Particularly, from a closer look at the spin-down Fermi surface of Cr$_3$Te$_4$, we notice an almost equal size of the hole ($k_F=0.12$ $au$) and electron($k_F=0.16$ $au$) pockets at the Brillouin zone center and zone corner, respectively [see Fig.~\ref{fig5}(g)]. This observation hints at a possibility of Fermi surface nesting as observed from the unconventional superconductors such as in the Fe-based~\cite{Chubukov2008, Mazin2008,Terashima2009} and Cu-based~\cite{Ruvalds1995,Valla2006} systems in their magnetic state. Moreover,  by the Sn intercalation, the size of the electron pocket increases at the zone corner, while the hole pockets are barely visible at the zone center [see Fig.~\ref{fig5}(h)]. Thus, the reduced Fermi surface nesting is demonstrated in the superconducting state of Sn$_{0.06}$Cr$_{3}$Te$_4$, again consistent with BaFe$_{2-x}$Co$_x$As$_2$ superconductor in which the highest T$_c$ was obtained when the size of the hole pockets at the zone center is significantly reduced~\cite{Thirupathaiah2010}.

\section{Conclusions}
In conclusion, we successfully induced superconductivity in the topological vdWs ferromagnetic Cr$_3$Te$_4$  by Sn intercalation at an onset transition temperature of $T_c\approx3.5$ K.  We conclude that Sn$_{0.06}$Cr$_{3}$Te$_4$ is a type-II superconductor with a lower critical field of $H_{c1}=58\pm4$ Oe and an upper critical field of $H_{c2}=209\pm10$ Oe. A jump in the specific heat noticed around the $T_c$ with a volume fraction of 33\% confirms the bulk superconductivity in Sn$_{0.06}$Cr$_{3}$Te$_4$. Spin-polarized DFT calculations on Cr$_3$Te$_4$ and Sn$_{0.06}$Cr$_{3}$Te$_4$  provide a better understanding on the orbital contributions near the Fermi level. Tuning of topological Hall effect originating from the skyrmion lattice is noticed with Sn intercalation in Cr$_3$Te$_4$.  Most importantly, for the first time, this study demonstrates superconductivity  in a skyrmion lattice, offering a new class of topological quantum materials.

\section{Acknowledgments}
A.B. acknowledges support from the Prime Minister's Research Fellowship (PMRF). S.G. acknowledges University Grants Commission (UGC),
India for the Ph.D. fellowship. A.N. thanks the startup grant of the Indian Institute of Science (SG/MHRD-19-0001) and DST-SERB (SRG/2020/000153). S.T. thanks the Science and Engineering Research Board (SERB), Department of Science and Technology (DST), India, for the financial support (Grant No.SRG/2020/000393).

\section{Data availability}
The raw and processed data required to reproduce these findings will be made available upon request to the corresponding authors.

\bibliographystyle{model1-num-names}
\bibliography{Cr3Te4_Sn_Clean}

\newpage
\section{Supplementary Information}

\begin{figure*}[htbp]
    \centering
     \includegraphics[width=0.7\linewidth]{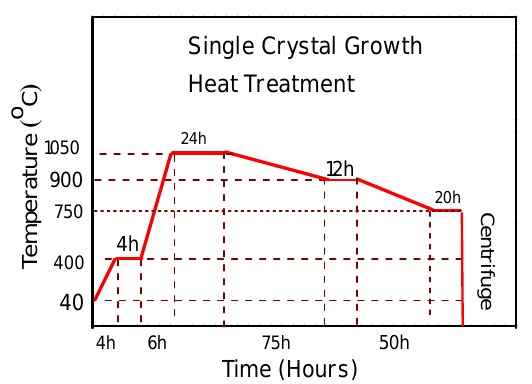}
    \caption{Heat treatment diagram of Sn$_{0.06}$Cr$_3$Te$_4$ single crystal growth.}
   \label{S6}
\end{figure*}

\begin{figure*}[htbp]
    \centering
     \includegraphics[width=0.7\linewidth]{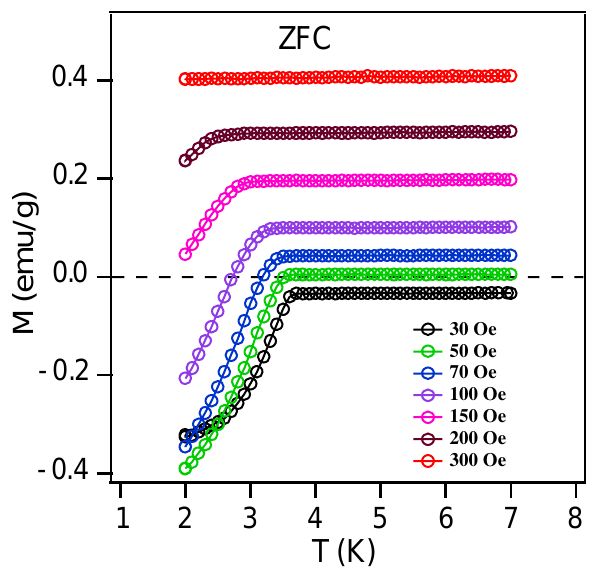}
    \caption{(a) Magnetization plotted as a function of temperature measured at various magnetic fields for $H\parallel \it{a}$ from Sn$_{0.06}$Cr$_3$Te$_4$. Clear Meissner effect with negative magnetization has been noticed below $T_c$ for the fields up to 100 Oe. Beyond 150 Oe, though a magnetization drop is noticed below $T_c$,  Meissner effect is absent. At 300 Oe, no drop in magnetization is noticed.}
   \label{S1}
\end{figure*}

\begin{figure*}[htbp]
   \centering
    \includegraphics[width=0.9\linewidth]{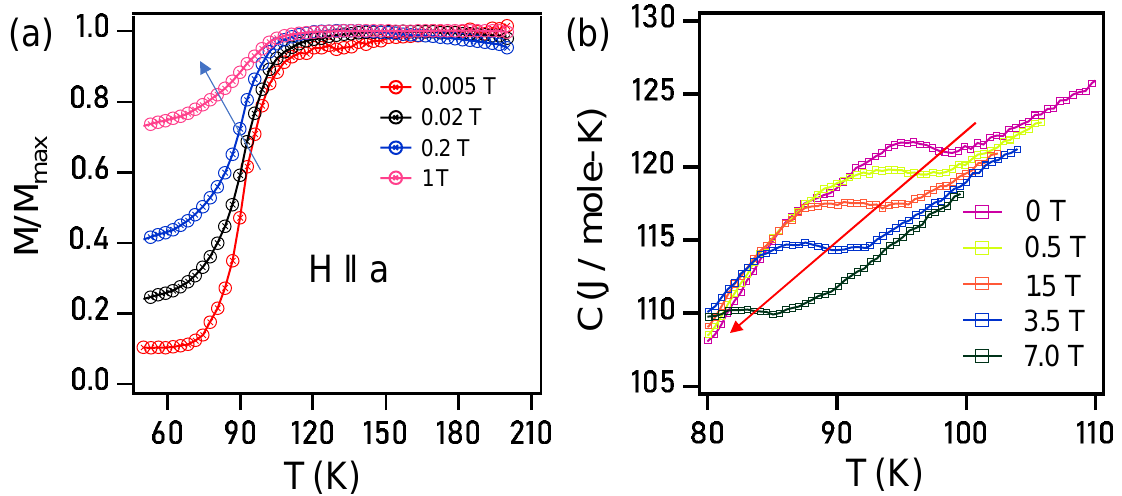}
   \caption{(a) Magnetization and (b) Specific heat plotted as a function of temperature, measured under various magnetic fields with $H\parallel \it{a}$ for Sn$_{0.06}$Cr$_3$Te$_4$. Both magnetization and specific heat data suggest that the magnetic transition found at 98 K is of the canted AFM-type as the magnetic transition temperature decreases with increasing the field.}
 \label{S8}
\end{figure*}


\begin{figure*}[htbp]
    \centering
    \includegraphics[width=0.8\linewidth]{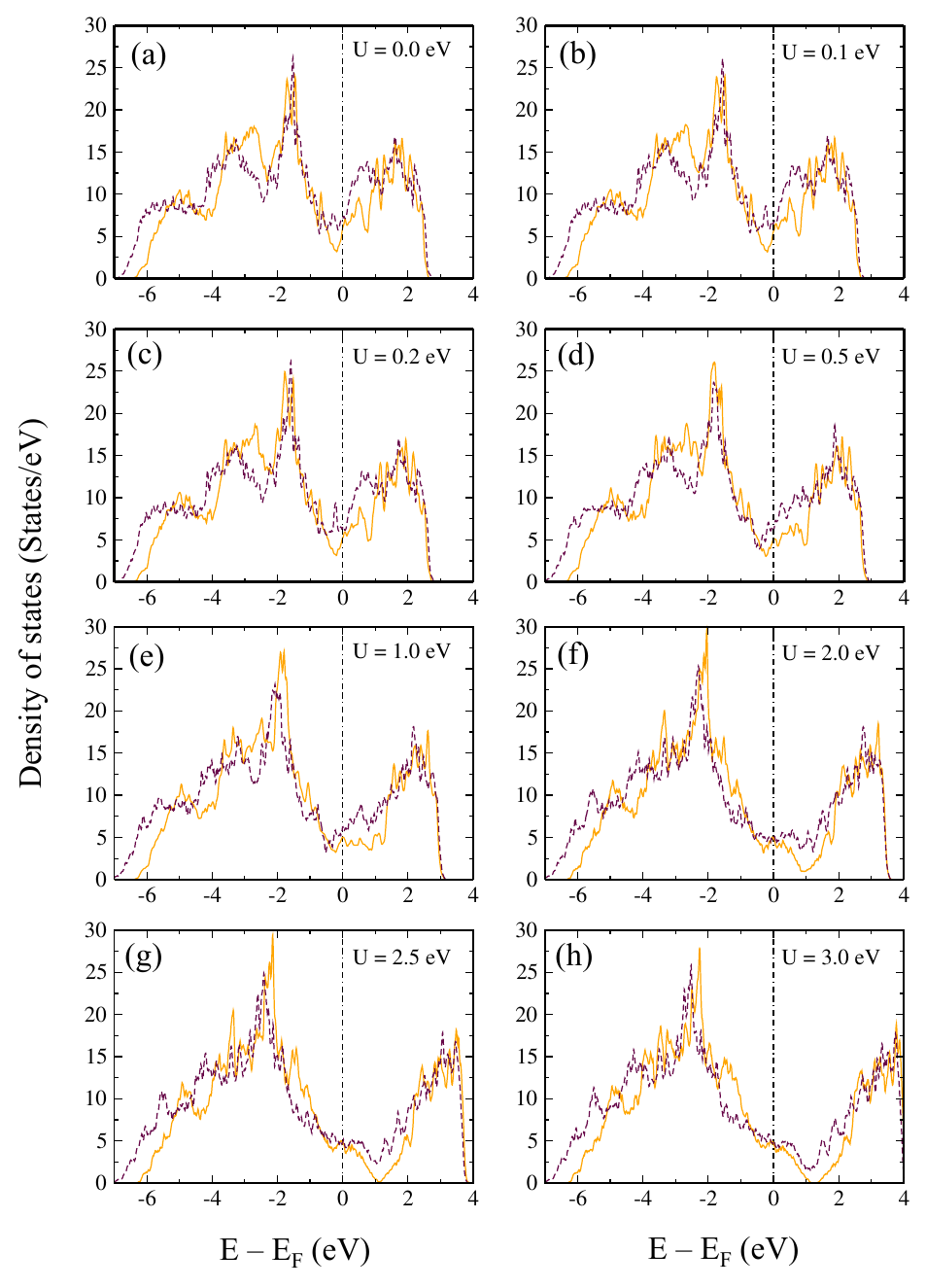}
    \caption{Density of states of parent Cr$_3$Te$_4$ (yellow solid line) and 6\% Sn intercalated Cr$_3$Te$_4$ (maroon dashed line) calculated at different values of the Hubbard (U) parameter.}
    \label{S2}
\end{figure*}

\begin{figure*}[htbp]
    \centering
    \includegraphics[width=\linewidth]{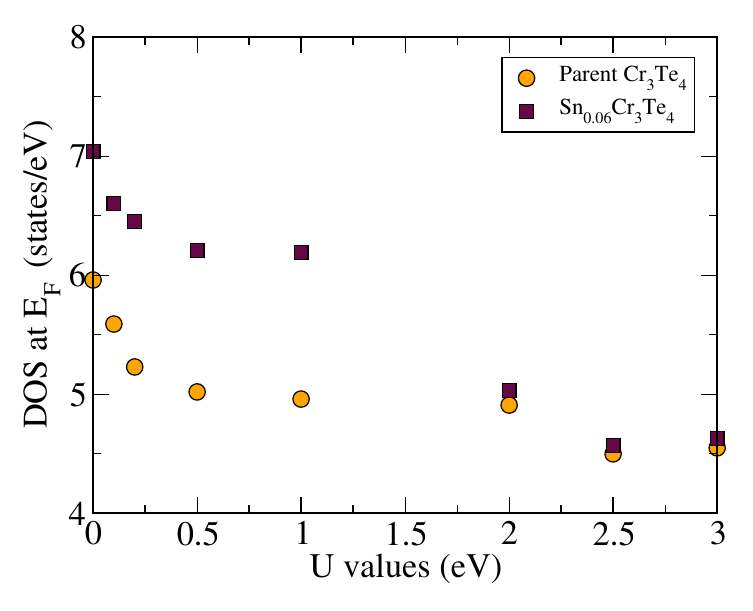}
    \caption{DOS at E$_F$ as a function of U. Yellow circles and maroon squares represent the parent Cr$_3$Te$_4$ and Sn$_{0.06}$Cr$_3$Te$_4$ systems, respectively.}
    \label{DOSatEF}
\end{figure*}

\begin{figure*}[htbp]
    \centering
    \includegraphics[width=\linewidth]{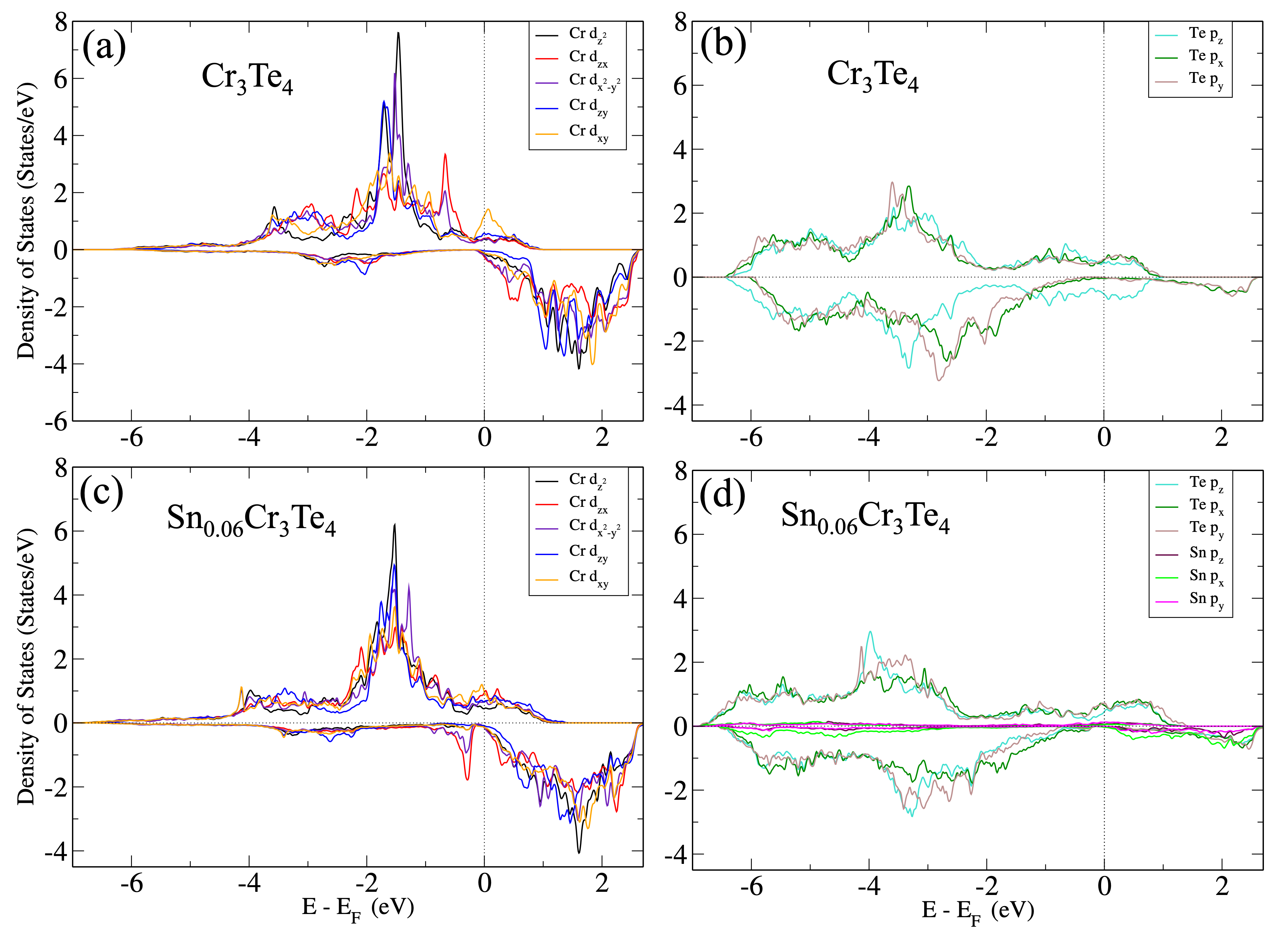}
    \caption{Projected density of states (PDOS) from parent Cr$_3$Te$_4$ for Cr-$d$ orbitals (a) and Te-$p$ orbitals (b). PDOS of Sn intercalated system for Cr-$d$ orbitals (c), and for $p$ orbitals of Te and Sn atoms (d). The negative values indicate down-spin states, while the positive values show the up-spin states.}
    \label{S4}
\end{figure*}

\end{document}